# Scanning Diamond NV Center Probes Compatible with Conventional AFM Technology.

Tony X. Zhou[1,2,‡], Rainer J. Stöhr[1,3,‡], Amir Yacoby[1,2,*]

[1]Department of Physics, Harvard University, 17 Oxford Street, Cambridge, Massachusetts 02138, USA.

[2]John A. Paulson School of Engineering and Applied Sciences, Harvard University, Cambridge, Massachusetts 02138, USA.

[3]3rd Institute of Physics, Research Center SCoPE and IQST, University of Stuttgart, 70569 Stuttgart, Germany

**Abstract:** Scanning probe microscopy using nitrogen vacancy (NV) centers in diamond has become a versatile tool with applications in physics, chemistry, life sciences and earth and planetary sciences. However, the fabrication of diamond scanning probes with high photon collection efficiency, NV centers with long coherence times and integrated radio frequency (RF) remains challenging due to the small physical dimensions of the probes and the complexity of the fabrication techniques. In this work, we present a simple and robust method to reliably fabricate probes that can be integrated with conventional quartz tuning fork based sensors as well as commercial silicon AFM cantilevers. For the first time, an integrated RF micro-antenna for NV center spin manipulation is directly fabricated onto the probe making the design versatile and compatible with virtually all AFM instruments. This integration marks a complete sensor package for NV center-based magnetometry, and opens up this newly developed scanning probe technique to the broader scientific community.

**Introduction**

Nanoscale magnetic sensors have become an integral part of contemporary condensed matter physics. Over the course of several decades, a variety of complementary techniques have been developed such as microscopy based on the magneto-optic Kerr effect (MOKE)[1], magnetic force microscopy (MFM)[2], Lorentz microscopy[3], and scanning-probe microscopy using superconducting quantum interference devices (SQUID)[4]. Each of these techniques has its own advantages and disadvantages when it comes to detection sensitivity, spatial resolution, bandwidth, and range of operating temperatures. Scanning-probe magnetometry using nitrogen vacancy (NV) centers in diamond is the latest addition to this family. Some of its key advantages are its high spatial resolution, and ultra-high sensitivity to magnetic field[5] while being suitable for room temperature studies and cryogenic applications alike. It was utilized to image a single electron spin at room temperature[6], superconducting vortexes[7,8], magnetic vortex states[9,10], hard drive domains[5], microwave current[11], magnetic domain walls[12–15], and skyrmions[16]. Therefore, NV center-based scanning-probe microscopy will contribute enormously to the broader communities in spintronics[17,18], chemistry[19–23], life sciences[24–30], and earth and planetary sciences[31,32].

Scanning NV center magnetometry started out using probes made of nano-diamonds glued to AFM tips[9,10,12–15,33–38]. In recent years, monolithic diamond nanopillars have been fabricated on thinned down diamond cantilevers to increase photon collection efficiency[5–8,11,39,40]. Single photon count rates of up to $1.4 \times 10^6$ per second[41] could be observed with $T_2$ coherence times typically around 30-90 μs [39]. However, fabrication and handling of monolithic diamond membranes as thin as 1-5 microns is challenging, making it difficult to manipulate and attach such micron-sized diamond cantilevers onto a scanning-probe platform.

In this work, we demonstrate a simple procedure to create diamond probes for scanning probe applications. Minimum fabrication steps are implemented to obtain large quantity of probes in parallel. In addition, the size of the probes is designed to be large enough to be compatible with commercial tipless AFM cantilevers. Additionally, we demonstrate the integration of a micro-antenna onto the AFM chip which delivers RF excitation to the NV center located inside a nanophotonic waveguide structure.

**Results**

**Design and fabrication of diamond probes**

Ultrapure electronic grade (100)-oriented CVD diamond substrates (13C natural abundance, Element Six) are cut and polished to be about 50 μm thick with about 1nm rms surface roughness. To remove polishing-induced defects and strain, one side of the substrate is etched by about 5 microns using oxygen RIE (Plasma-Therm Versaline ICP–RIE). Subsequently, NV centers are created on that side using implantation of nitrogen 15 with an implantation energy of 6 keV followed by thermal annealing[42]. Figure 1a summarizes the main steps of the fabrication of detachable diamond cubes which are later used as scanning magnetometry sensors.

First, diamond nanopillars with a diameter of roughly 350 nm are fabricated on the nitrogen implanted side of the diamond following a recipe described in the supplemental information[42]. The length of the nanopillar can be adjusted by the etching time and was chosen to be roughly 3.5μm. During nitrogen implantation, the radiation dose was chosen such that each nanopillar hosts on average a single NV center. Subsequently, the shape of the diamond probes is lithographically defined on the other surface of the substrate using photoresist. For this, a photolithographic mask is aligned with respect to the diamond nanopillars visible on the downfacing side of the substrate such that the nanopillar is close to the front apex of the probe (see Figure 2d). The probe dimensions are chosen to be 125μm in length, 50μm in width, and 50μm in thickness, resembling an elongated cube shape. After exposing and developing the photoresist, a 400nm thick layer of titanium is thermally evaporated on the structured diamond surface. After lift-off of the photoresist mask, this layer serves as an etch mask during the subsequent dry etching step. In this step, the titanium side of the sample is exposed to oxygen RIE which etches all unprotected regions of the diamond through its entire thickness. After removal of the residual titanium layer, this results in an array of individual diamond probes which are attached by tiny joints to the substrate frame as shown in Figure 1b. Finally, the entire diamond substrate is cleaned in a boiling acid mixture consisting of equal parts of sulfuric, nitric, and perchloric acid to remove contaminants from fabrication and to oxygen terminate the surface.

Diamond fabrication is hard partly due to many acid cleans. Starting from nanopillar fabrication, the number of acid clean is minimized to only 1 at the final step. When designing the photolithographic mask and also during the optimization of the plasma etching recipe, special attention is given to the size and shape of the joints. They are designed to be strong enough not to break during wet chemical treatments (acid clean) of the entire structure, yet weak enough to allow the diamond cubes to be released when attaching the probes to a scanning platform as described below. The outlined fabrication process yields roughly 52 probes on a 2x4mm² substrate.

**Characterization of diamond probes**

Before further processing, each diamond probe is characterized in a homebuilt confocal microscopy setup[42]. First, optically detected magnetic resonance (ODMR) technique[43,44] is used to identify all pillars hosting at least one NV center (see Figure 1c). Nanopillars hosting only a single NV center are further distinguished by performing second-order autocorrelation measurements (see Figure 1d). The brightness of these NV centers is further determined by measuring their saturation count rate and saturation laser power (see Figure 1e). On average, out of 52 probes on one substrate, 15 show strong, photostable, single NV center emission with a count rate of 200-500×10$^3$ per second and are therefore considered usable for further scanning probe application. Among these usable probes, the average coherence time T$_2$ is found to be 61µs (see supplemental information for histogram). The T$_2$ time of a specific NV center can vary strongly based on its location inside the nano-pillar and its electric and magnetic environment as well as crystal strain. It has been well studied that paramagnetic spins on diamond surface and $^{13}$C nuclear spins are main sources of decoherence for shallow NVs inside nano-pillars[45].

**Integration of probes onto AFM platforms**

Two of the most common AFM feedback platforms are optical beam deflection and quartz tuning fork[46]. Most commercial AFM instruments rely on the former due to its compatibility with quickly exchangeable and standardized silicon cantilevers. However, many homebuilt scanning-probe setups have been using predominantly conventional quartz tuning fork based sensors due to their simple implementation and compatibility with low temperature conditions[47]. In the following, we show how the diamond sensors described above can be reliably integrated into both of these platforms using very basic equipment and simple procedures.

For the case of AFM beam deflection sensors, the probes are directly glued to tipless AFM cantilevers (see Figure 2c). For this, a small drop of UV curable adhesive is applied to the top surface of a diamond cube. Under an optical stereo microscope, the AFM cantilever is then mounted to a manual translation stage and positioned on top of a diamond cube touching the adhesive drop. After curing the glue under UV light, the diamond probe is detached from the substrate frame by breaking the weak joints using a sharp tungsten tip mounted to a separate manual translation stage (see supplemental info for more details[42]). Figure 2e shows an example of a diamond cube glued to an AFM cantilever, which can then be further used in a scanning geometry using optical beam bounce methods (illustrated in Figure 2f). For the case of tuning fork based sensors, a diamond cube is first glued to a pulled quartz rod (Figure 2a) following a similar procedure as in the case of AFM cantilevers. The quartz rod is then attached to one prong of the tuning fork (see Figure 2b). It is worth noting, that the mounting techniques described here do not require the use of any sophisticated equipment such as focused ion beam (FIB) assisted gluing and more time-consuming recipes involving nano-manipulation of the diamond slab. Here, the increased size of the diamond probe mitigates these complications without compromising the optical performance of the probe or the spin properties of the embedded NV center. In addition, diamond cubes of 50μm in size are ideally suited for use with commercial AFM cantilevers that are typically between 30μm and 70μm in width.

Another experimental aspect of scanning NV center magnetometry is the need for RF signal that drives and controls the NV center. Traditionally, this is achieved using an RF waveguide that is fabricated onto the sample substrate or by introducing a small antenna loop in between the sample and the objective lens using additional translation stages. The former requires additional fabrication steps during sample fabrication and the latter results in increased experimental complexity. Therefore, the ability to integrate RF components onto the probe is desirable in particular for cryogenic applications. Figures 3a and 3d show a simple RF micro-antenna integrated right above the diamond probe attached to a silicon cantilever. The micro-antenna is wire-bonded to the silicon AFM chip and bent by a tungsten tip mounted to a linear translation stage to be positioned in proximity to the diamond probe. A 500nm thick oxide layer is initially grown on the silicon chip in order to provide electrical insulation between the bond pads. The bond pads are then connected to an RF source in order to apply the RF signals in close vicinity of the NV center. In this way, Rabi oscillations of the NV center can be driven over a wide frequency range at micro-antenna with an input power of 30 dBm (Figure 3b). Rabi frequencies as high as 4.8 MHz can be reached at an input power of 35 dBm (Figure 3e). The decreasing Rabi rate at higher frequencies is consistent with the microwave transmission measured using a network analyzer (see supplemental info[42]). This demonstration strongly encourages further engineering to fabricate a silicon AFM chip such that an RF stripline can be lithographically patterned nearby the cantilever.

**Scanning probe microscopy demonstration**

We demonstrate the functionality of the diamond probe by performing independently an AFM measurement as well as a magnetic field scan using the probes. The AFM height measurements is performed using Bruker Bioscope Catalyst in contact mode to map out the topography of a calibration sample consisting of about 178nm deep square pits in 10μm pitch. Figure 3f correctly matches the real shape of the pits as confirmed by AFM measurements using sharp commercial AFM tips. Given the rather large footprint of the diamond probe in comparison to the nanopillar height, care must be taken to not tilt the probe by more than 8 degrees relative to the sample surface. A tilt in excess of 8 degrees causes the edge of the probe rather than the nanopillar to touch the sample surface. Figure 3c shows the mechanical resonance spectrum of the cantilever with and without the diamond probe. The fundamental mode of the tipless cantilever is detected at 17 kHz, and it is observed to be shifted to 2.9 kHz with the diamond probe attached due to the additional mass added to the cantilever. However, the quality factor of the resonance is not affected by the added mass.

Finally, scanning NV center magnetometry was performed in a home built confocal microscope. For this, a diamond probe glued to a quartz rod is mounted on a quartz tuning fork so that the nanopillar containing a single NV center can engage the surface of a quartz substrate where a coplanar waveguide is fabricated and wire bonded to an RF source. AC magnetic fields at radio frequencies are generated within the gap of the waveguide. The sample is mounted on a 3D piezoelectric nano-positioning stage. Using the AFM feedback mode, the sample is approached to the diamond probe until the nanopillar is in contact with the sample surface. An AFM topography image (see Figure 4a) inside the gap of the waveguide is taken by moving the sample while maintaining the nanopillar in contact with the surface. Within the gap, the RF excitation generates Rabi oscillations of the NV center electron spin. By measuring the Rabi frequency, the magnetic field is quantitatively measured. A line scan of AC magnetic field is demonstrated in Figure 4b. The field profile is modeled by considering two counter-propagating currents based on Appel *et al*[11]. By solving Maxwell's equations, the magnetic field profile along the NV center axis is obtained (see Figure 4b). As expected, the local peaks of the magnetic field are located at the edges of the central conductor and the ground plane. The NV center to sample distance is extracted to be 20nm comparable to previous report[5] using the same implantation energy. This distance can be controlled in two ways. First, one can retract the sample relative to the diamond tip by conducting experiments without continuous AFM feedback. Second, one can change the implantation energy of nitrogen ions to control the depth of NV centers relative to nano-pillar end (see Supplemental information for NV center depth at various implantation energies). The latter ultimately sets the minimum NV center to sample distance.

**Conclusions**

In summary, a novel, robust, and simple fabrication technique for diamond probes hosting single NV centers in a monolithic diamond structure is presented. The probes are easily integrated into the two most common AFM feedback platforms, conventional quartz tuning fork based sensors and optical beam deflection method. Furthermore, the integration of a micro-antenna into the probe sensor for RF excitation of the NV center is demonstrated. The performance of the NV center nanophotonic structure is assessed by second-order autocorrelation measurements, as well as saturation measurements and the detection of the NV center ODMR signal. The performance of the entire sensor assembly is demonstrated by measuring surface topographies as well as spatially resolved magnetic field maps. We believe, this will allow scanning NV center magnetometry to be more accessible to a wider scientific community. Furthermore, all of the above can be extended to other species of color centers in diamond for a variety of scanning probe applications.

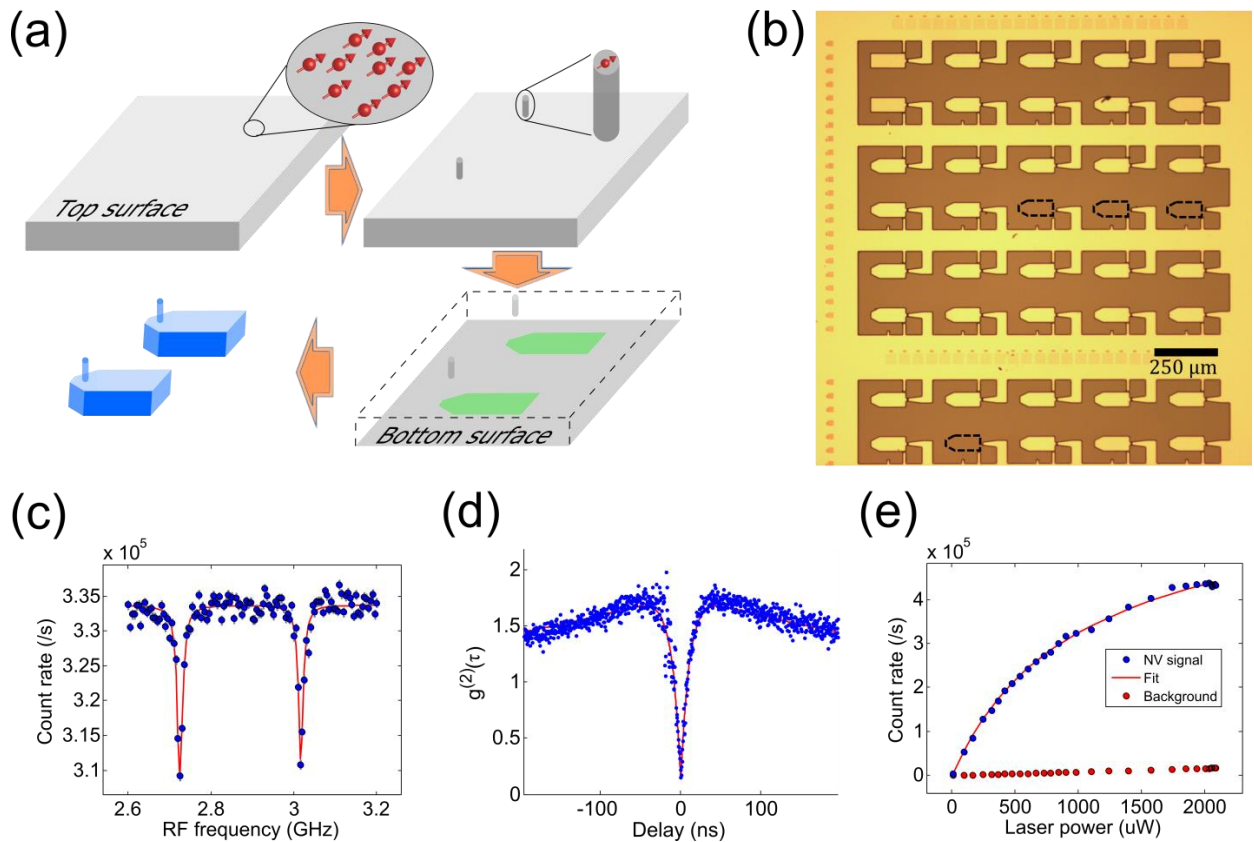

**Figure 1**. Fabrication and characterization of diamond probes. (a) Schematic of the fabrication procedure starting from a diamond substrate with shallow-implanted NV centers followed by nanopillar formation and diamond probe fabrication. (b) Optical image of diamond substrate showing diamond probes after fabrication. Characterization of NV centers inside diamond probe is performed by (c) optically detected magnetic resonance technique (ODMR) at bias field 52 Gauss, (d) second-order autocorrelation, and (e) saturation of photoluminescence.

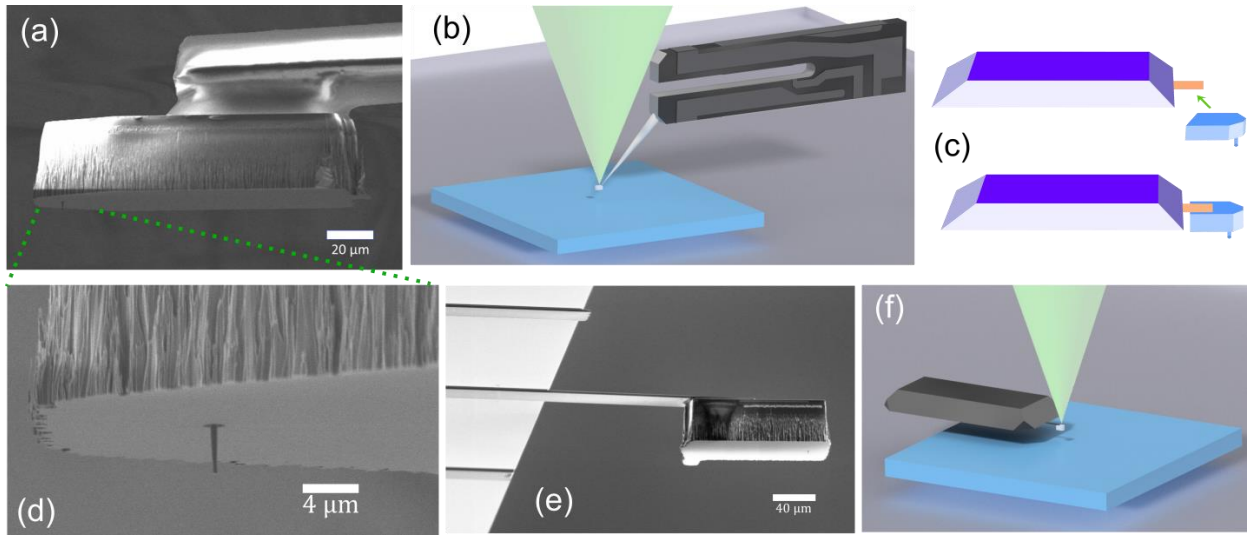

**Figure 2**. Integration of the diamond probes into the two most common AFM feedback platforms: (a) SEM image of a diamond probe glued onto a quartz rod mounted to a tuning fork. (b) Shows a schematic of the sensor geometry. (c) Illustration of diamond probe integration to silicon cantilever AFM tips. (d) SEM image of diamond nanopillar located near the left edge of diamond probe (e) SEM image of a diamond probe glued onto a commercial tipless AFM cantilever. (f) Shows the geometry used for optical beam deflection feedback.

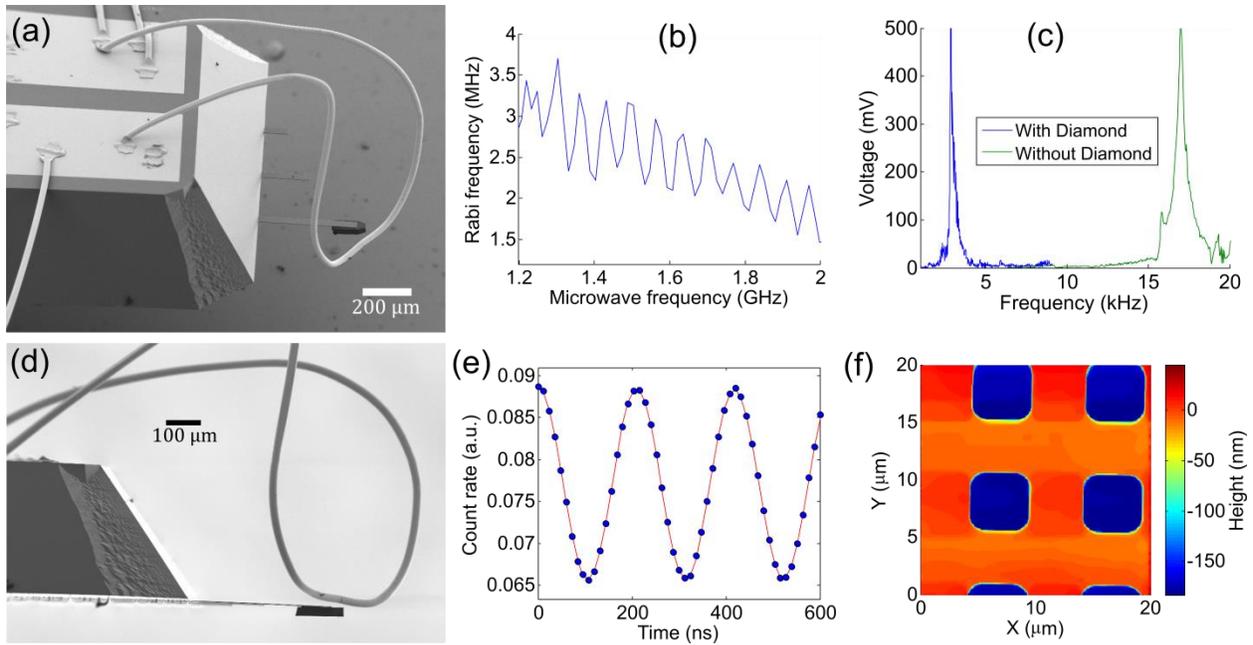

**Figure 3**. Micro-antenna integration onto silicon cantilever AFM tips for RF excitation of NV center: (a) SEM image of gold antenna placed near a diamond probe glued on silicon cantilever (b) Measurement of Rabi frequency at different microwave frequencies at micro-antenna with an input power of 30 dBm. (c) Cantilever resonance before and after mounting diamond probe, (d) SEM image of probe in (a) at side angle. (e) Rabi frequency of 4.8 MHz can be observed at 1.75 GHz at 400 Gauss with micro-antenna input power at 35 dBm. (f) AFM height measurements of a calibration sample with diamond probe mounted on silicon cantilever.

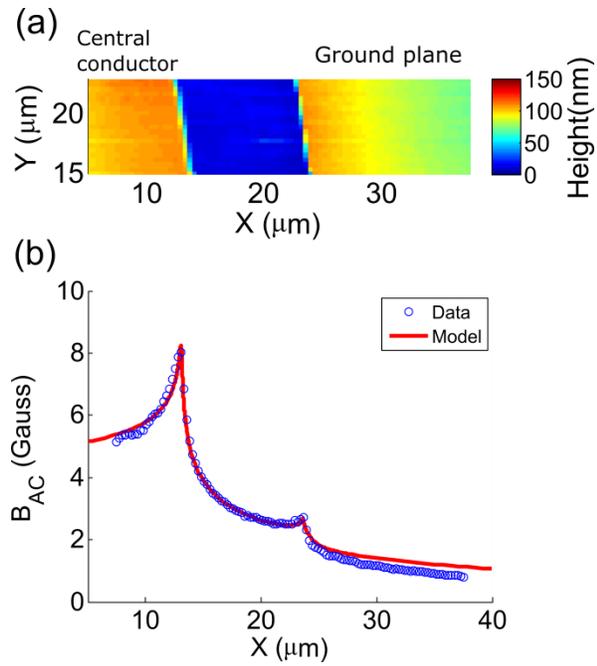

**Figure 4**. Scanning NV center magnetometry performed on coplanar waveguide: (a) Topography of coplanar wave guide gap. (b) AC magnetic field measurement across the gap at 1.6 GHz frequency of RF and 450 Gauss bias field.

**Corresponding Author**

*Address: LISE sixth floor, 11 Oxford Street, Cambridge,
Massachusetts 02138. E-mail: yacoby@physics.harvard.edu.

**Author Contributions**

‡These authors contributed equally to this work

**Funding Sources**

This work is supported by the Gordon and Betty Moore Foundation's EPiQS Initiative through Grant GBMF4531. A.Y. is also partly supported by the ARO grant W911NF-17-1-0023.

**Acknowledgments:**

Sample fabrication was performed at the Center for Nanoscale Systems (CNS), a member of the National Nanotechnology Coordinated Infrastructure (NNCI), which is supported by the National Science Foundation under NSF award no. ECCS - 1541959. CNS is part of Harvard University. We thank Mathew Markham and Element Six (UK) for providing diamond samples. We thank Bruker nano surfaces division for discussion and technical support, and Weijie Wang for providing silicon cantilevers. We also thank Ronald Walsworth and Matthew Turner for annealing diamonds, and S. Ali Momenzadeh, Mike Burek, and Marc Warner for fruitful discussions. During revision of the published manuscript, we thank Patrick Maletinsky and Patrick Appel for useful discussions.